# Airborne Base Stations for Emergency and Temporary Events

Alvaro Valcarce[1], Tinku Rasheed[2], Karina Gomez[2], Sithamparanathan Kandeepan[3], Laurent Reynaud[4], Romain Hermenier[5], Andrea Munari[5], Mihael Mohorcic[6], Miha Smolnikar[6], and Isabelle Bucaille[7]

[1] TriaGnoSys GmbH, Wessling, Germany
[2] Create-Net, Trento, Italy
[3] RMIT University, Melbourne, Australia
[4] Orange, Lannion, France
[5] DLR (German Aerospace Center), Wessling, Germany
[6] Jozef Stefan Institute (JSI), Ljubljana, Slovenia
[7] Thales Communications & Security, Paris, France

**Abstract.** This paper introduces a rapidly deployable wireless network based on Low Altitude Platforms and portable land units to support disaster-relief activities, and to extend capacity during temporary mass events. The system integrates an amalgam of radio technologies such as LTE, WLAN and TETRA to provide heterogeneous communications in the deployment location. Cognitive radio is used for autonomous network configuration. Sensor networks monitor the environment in real-time during relief activities and provide distributed spectrum sensing capacities. Finally, remote communications are supported via S-band satellite links.

**Key words:** Emergency, LTE, sensor networks, altitude platform.

## 1 Introduction

The rapid growth of bandwidth-hungry communication applications has put network operators under pressure to exploit the radio spectrum as efficiently as possible. Furthermore, recent events have shown that in the aftermath of an unexpected event, communication infrastructures play an important role in supporting critical services such as emergency recovery operations, infrastructure restoration, post-disaster surveillance, etc [1].

In this context, this paper considers the provision of rapidly deployable and resilient networks to provide Broadband access in large coverage areas. Current mission critical systems, including Public Protection and Disaster Relief (PPDR) communication systems, are limited in terms of network capacity and coverage. They are not designed for or suitable to address large scale emergency communication needs in a disaster aftermath. PPDR systems are also limited by interoperability barriers, the technological gap with commercial technologies and evolving standards. Further, the terminals of first responders are getting smarter thanks to the introduction of new applications with integrated sensors and to the availability of multi-mode transceivers for supporting for example video conferencing, near-real-time video streaming and push-to-talk. Such improvements lead to an increase in capacity and energy demands at the terminals. Moreover, some events with large aggregation of professional and consumer users such as e.g. big sporting events or concerts require high capacity and/or dedicated coverage that the legacy terrestrial network infrastructure cannot rapidly provide.

Airborne communication networks have been recently studied for the provision of wireless communication services and have continually attracted interest from government, industry and academia [2]. Most of the original efforts focused on long endurance High Altitude Platforms (HAPs) [3] operating at altitudes of about 17-25 km. However, other types of airborne platform including aerostats and aerodynes, have been recently developed to fly at altitudes between a few dozen and a thousand meters. Those platforms, gathered under the denomination of Low Altitude Platforms (LAPs), are increasingly believed to complement conventional satellite or terrestrial telecommunications infrastructure. For example, the Deployable Aerial Communication Architecture (DACA) architecture proposed by the FCC in the US explores the feasibility of LAPs to be employed during emergency situations to restore critical communications [4].

In this paper, a holistic and rapidly deployable mobile network is presented. This is composed of a LAP-based airborne communications segment combined with a satcom-enabled Portable Land Rapid Deployment Unit (PLRDU). The solution, developed within the framework of the ABSOLUTE project [5], attempts to demonstrate the high-capacity, low-latency and large coverage capabilities of LTE-A for the provision of broadband emergency communications. In addition, the system integrates heterogeneous radio access technologies (i.e. LTE-A, TETRA, WLAN, S-band satellite links, etc) to fulfill coverage, data rate and latency requirements. Finally, technical challenges and methods for autonomous network reconfiguration in such scenarios are discussed.

## 2 Scenarios & Use Cases

The architecture described here is applicable to the event types described next.

### 2.1 Public Safety Communications

In the wake of a disaster, multiple PPDR agencies need to operate collaboratively on the rescue site using reliable and interoperable communication systems [6]. In this context, prominent services include first responder communications, critical infrastructure restoration support systems, post-disaster surveillance, medical service networks, etc. However, preexisting networks may have been damaged or destroyed. For instance, base stations from cellular networks may have been hit by an earthquake or tsunami, as well as affected by power outages induced by multiple causes such as severe weather events, including floods and hurricanes [7].

Some disasters or emergency events may, to some extent, be anticipated. However, the degree of anticipation varies greatly, and generally remains very limited [8]. In any case, first responders rely on communication devices, equipped with multiple sensors and heterogeneous transceivers to support increasingly bandwidth-demanding applications, including real-time video streaming or exchanges of large amounts of data (e.g. high resolution images, environmental or medical monitoring). This puts a strong demand on the PPDR community for reliable and scalable communication infrastructures to provide network coverage, low delay and high capacity, as well as interoperability with legacy radio technologies. In the proposed system architecture, typical public safety applications can be supported such as Half-duplex video conferencing, near-real-time video streaming, bulk file transfer, e–mail, web, LTE Push-to-talk and VoIP [9].

### 2.2 Capacity Enhancements During Temporary Events

In recent years, mobile network access has grown in coverage and capacity, especially in densely populated regions. However, in less populated areas, access to high-speed mobile networks did not follow this trend. In fact, Mobile Network Operators (MNOs) now face the task of defining economically viable network architectures for the deployment of high-speed services in suburban or rural zones. Such architectures are not necessarily permanent and could be deployed only during demand peaks.

As a result, the organizers of mass events to take place in areas with limited network capacity need affordable solutions to provide connectivity to the temporary users, such as attendees and the organizers themselves. Such temporary service can be deployed to accommodate different events including outdoor gatherings, conference and seminars, construction sites, festivals and sporting events [10]. Communication solutions for these cases could lean upon temporary networks, such as the LAP-based one proposed in this paper, thus replacing or complementing preexisting network infrastructure.

## 3 Architecture

Figure 1 illustrates the proposed deployment architecture to provide emergency and temporary communications. This network needs to be resilient, supportive of broadband applications and secure. In addition, network service shall be tailored to the type of disaster or temporary event. Moreover, the architecture must support rapid configuration and deployment of the broadband network. This is achieved through the design of two tightly interconnected segments: an air segment, which consists of LAPs, and a terrestrial segment, enabled by multiple cooperating terrestrial equipments.

Several radio access technologies can be offered from the LAP. These shall be chosen based on the required coverage and capacity, as long as LAP payload constraints are respected (i.e. size, power consumption and weight, etc). Recent approaches were based on the use of IEEE 802.11 access points and on femtocells [11] to support communications with terrestrial User Equipments (UEs) and surviving Base Stations (BSs). In this project [5], the design and deployment of low-complexity eNodeBs is analyzed. In particular, the so-called Airborne eNodeB (AeNB) is embedded in each LAP's communications payload, with includes inter-LAP links, thus supporting reliable communications for emergency or temporary events. The air segment also supports dynamic spectrum access, therefore to gain awareness of existing terrestrial networks and provide seamless network reconfigurability and scalability. The terrestrial segment consists of nodes such as the PLRDUs and the Multimode User Equipments (MMUEs), which are explained in the next section.

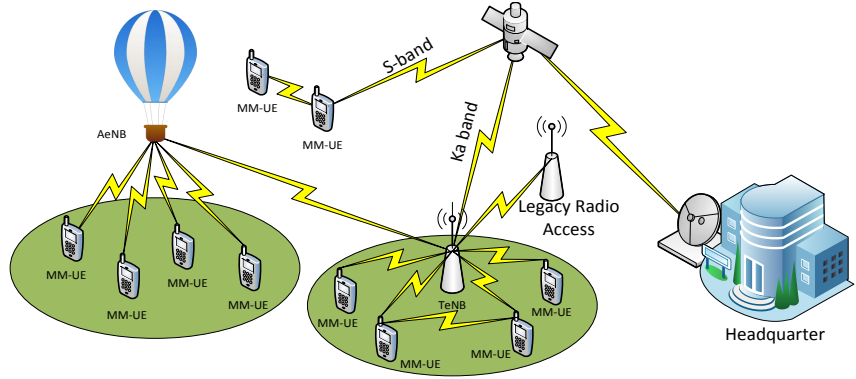

**Fig. 1.** Rapid deployment architecture for broadband access provision.

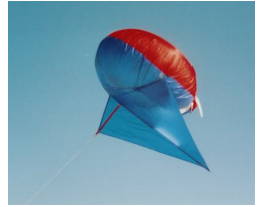
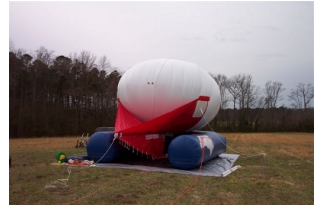

(a) Airborne Helikite.    (b) Helikite resting on a Helibase.

**Fig. 2.** Low Altitude Platform.

## 4 Subsystems

### 4.1 Low Altitude Platform (LAP)

The capabilities of the various interoperable nodes, which is explained next, turns the proposed network into a system that is more than the sum of its parts. As opposed to HAPs [3], a LAP is an airborne system capable of lifting communications equipment off the ground to heights between 300 m and 4 km. In the architecture described in this paper, the LAP is a lighter-than-air kite manufactured by Helikites [12] and known as a *Helikite* (see Figure 2(a)). The Helikite is an aerostat, thus relying on aerostatic lift to achieve buoyancy. It uses a helium-filled balloon to provide lift power, while the more aerodynamic kite provides stability against wind-caused drift. Additionally, the Helikite is tethered, which provides stability in changing winds. The Helikite has a lifting capacity of 10 kg and it can remain airborne for several days. However, climatological conditions may reduce its lifting capacity as well as its time aloft. When on the ground, the Helikite rests on an inflatable platform known as the *Helibase* (see Figure 2(b)) to prevent accidental punctures and to facilitate safe and rapid deployments.

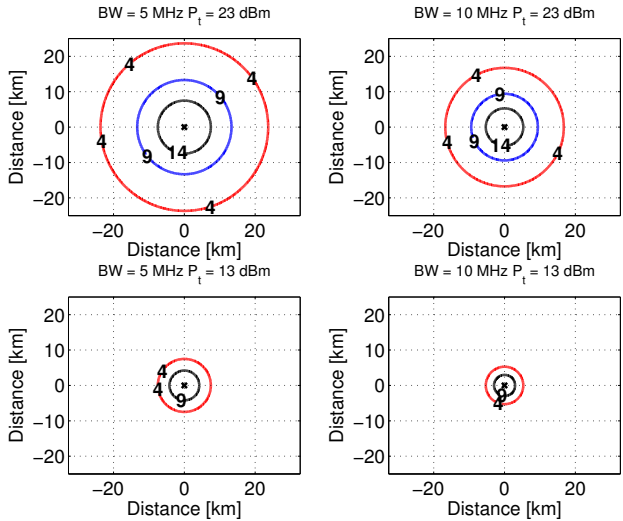

**Fig. 3.** Ground-level downlink SNR (dB) contours for an airborne LTE cell at 795.5 MHz and 300 m height. System parameters for these estimations are given in Table 1.

When lifting an eNodeB, RF signals overcome most ground-level obstacles such as trees, buildings, hills, etc. This enables nearly all UEs to enjoy a line-of-sight (LOS) to the AeNB, thus increasing the coverage area. For example, considering LOS free space propagation, Figure 3 shows the cell sizes than can be achieved from an AeNB in the absence of cochannel interference. This is illustrated with SNR isolines, where the SNR is calculated in dB as

$$SNR = P_r - N_{th} - F_{UE}, \qquad (1)$$

being $P_r$ the DL power received by the UE and calculated as

$$P_r = P_t + G_t - L_{fs} + G_{UE} - F. \qquad (2)$$

Similarly, $N_{th}$ is the thermal noise for a given system bandwidth $BW$ in Hz at the standard temperature $T = 293.15$ K and calculated as $N_{th} = 10*\log_{10}(k_B \cdot T \cdot BW)$, where $k_B$ is Boltzmann's constant in joules per kelvin. Furthermore, $F_{UE}$ is the noise figure of the UE, $P_t$ is the AeNB transmit power, $G_t$ the antenna gain of the AeNB, $L_{fs}$ the free space RF propagation losses, $G_{UE}$ the UE antenna gain and $F$ a fading margin. Finally, the calculations of Figure 3 use the values summarized in Table 1 and inspired by [13] for downlink LTE link budgets.

Table 1. System Parameters.

| Parameter | Fading margin | AeNB antenna gain | UE noise figure | UE antenna gain |
|---|---|---|---|---|
| Symbol | $F$ | $G_t$ | $F_{UE}$ | $G_{UE}$ |
| Value | 4 dB | 3 dBi | 7 dB | 0 dBi |

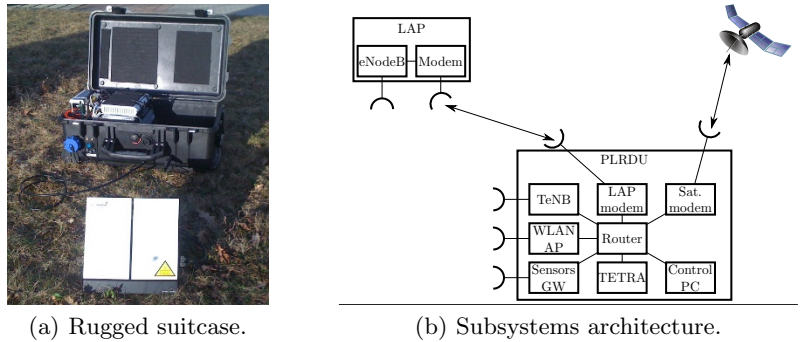

(a) Rugged suitcase.    (b) Subsystems architecture.

Fig. 4. Portable Land Rapid Deployment Unit.

### 4.2 Portable Land Rapid Deployment Unit (PLRDU)

The PLRDU is a standalone and self-sufficient communications platform. It takes the form of a rugged suitcase (see Figure 4(a)) to accommodate systems such as a WLAN access point, a Terrestrial eNodeB (TeNB), a sensors gateway, an IP router, an interface to a TETRA system and a Ka band satellite modem. Additionally, the PLRDU also includes subsystems that support its main role as a communications platform. These are: batteries, power supply and a PC that controls all of the PLRDU functions. The PLRDU is designed to be easily transported to locations where a communications network is required. This enables wireless transmissions in the destination area, where local communications are piped through a router and remote communications through a satellite link.

Despite its ability for self sustainment and independence from the other subsystems, the PLRDU plays an integrating routing role within the larger system architecture. This is achieved through the Ka band satellite modem and the local IP router, that provide remote connectivity to the systems on board of the LAP. IP traffic is exchanged between the PLRDU and the LAP through means such as cabling (through the Helikite tether), WLAN or even optical radio links. The "LAP modem" in Figure 4(b) illustrates the heterogeneous nature of this link. Using the PLRDU to provide Internet and PSTN connectivity through satellite backhauling avoids the need for satellite modems on the LAP, whose lifting capacity is limited. Besides, orienting a satellite antenna from a balloon platform is more challenging than relaying the LAP traffic to the ground-based PLRDU.

### 4.3 Multimode User Equipment

As illustrated in Figure 1, MMUEs support LTE and TETRA radio interfaces for communicating within TeNB or AeNB coverage. In addition, direct reception of narrowband broadcast satellite services in S-band are supported to guarantee the resilience of the network infrastructure. To meet the requirements of emergency recovery situations, the interoperability between infrastructure and proximity communications exploits the dynamic service discovery as well as the imposed energy constraints to maximize the overall system lifetime.

Apart from supporting the ubiquitous multi-mode high speed wireless broadband coverage, MMUEs also feature functionalities of sensor data fusion, localization and identification. In this respect sensors embedded in the handheld UE provide to the user and other members of the rescue team with context about the environment (e.g. temperature, humidity, luminosity, concentration of toxic or hazardous gases, etc). This information is mapped to the relative and/or absolute MMUE location (by means of GPS or ad hoc established infrastructure for localization exploiting triangulation principles), and movement (e.g. by means of accelerometer, electronic compass and gyroscope sensors). With respect to identification the capabilities of the system allow users to combine different parts in a virtual network. The main interfaces used for these purposes are RFID, NFC or special biometric sensors.

Wireless sensors in the form of body area networks can also be used in emergency scenarios to monitor the status of first responders. In such case a mobile sensor node can be used as a personal gateway between the UE and the body sensors. These can be used for monitoring heart beats, blood pressure, breathing, sweating, stress, presence and concentration of various gases, ambient temperature, spot temperature, radiation, etc. All of these can be embedded in the first responder's equipment or clothing.

Various radio interfaces in the MMUE can also be used for distributed monitoring of the radio environment and complemented by externally distributed spectrum sensing nodes (see section 5). This could be used to build radio environment maps and spectrum occupancy tables.

### 4.4 Sensors Network

In a disaster-hit area, first responders rely on previously existing reports (if available) and on their own senses. In order to support planning activities and actions in the field, ad-hoc self-organizing sensor networks could play a role in monitoring the environmental conditions. Such networks would diminish risks for members of the rescue teams and gather real-time data to monitor the evolution of the disaster-relief activities.

The need for monitoring critical infrastructure and the operating environment where first responders work is addressed in this project by the development of scenario-tailored sensor modules. For example, specific modules are foreseen for flooding scenarios where water levels and flows need to be monitored. Similarly, earthquake scenarios would require of aftershock monitoring, sensing of infrastructure cracks. In nearly all disaster-hit scenarios, environmental conditions (weather, gas presence, etc) need to be monitored. Additionally, wireless sensors can be used for distributed spectrum sensing, which would support the constructions of environmental radio maps. This would support the set up of emergency communication systems in interference-free frequency bands.

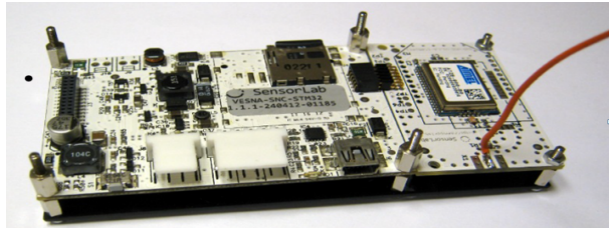

**Fig. 5.** VESNA wireless sensor gateway to be integrated into the PLRDU.

The modules described above take the form of application specific expansion boards for sensor nodes based on the VESNA platform [14]. These sensor nodes are equipped with a radio interface, a battery and if necessary an energy harvesting module. They set up ad-hoc sensor networks and connect them to the sensor gateway in the PLRDU (see Section 4.2). Scenario-specific sensor modules thus represent the building blocks of the various scenario-specific VESNA deployments. Figure 5 illustrates the core module and radio boards of the VESNA gateway, which is to be integrated in the PLRDU.

It is important to note that this paper aims to present the overall concept of supporting emergency and temporary events in which also sensor networks play an important role, on one hand serving as environmental sensors supporting safe operation of rescue teams, and on the other hand when used as spectrum sniffers supporting operation of cognitive radio enabled communication equipment. Clearly, both uses will also affect the system capacity, the extent of this affect depending also on the number and type of sensors, and need to be taken into account in future work when we will look deeper into capacity issue.

### 4.5 Satellite Backhauling

One key feature of the presented architecture is its capacity for seamlessly resorting to satellite backhauling when necessary. Such an approach brings multi-fold benefits. On the one hand, the most cost effective User Plane (UP) traffic backhauling means can be dynamically chosen when the terrestrial communication infrastructure is unavailable (e.g. thus enabling communications with the safety operations control centre in case of disaster), or insufficient to cover traffic peaks(e.g. satisfying increased internet network demand during temporary events). On the other hand, the routing data through satellite links paves the way for the investigation and design of solutions which dynamically switch the backhauling strategy so as to maximize network capacity and performance.

The IP-based satellite link will provide UP traffic backhauling connectivity between the TeNBs, the Internet and the Public Switched Telephone Network (PSTN). This is achieved through a geostationary satellite operating in Ka band. The availability of satellite communications can also be exploited to establish Internet connections for the users. In this perspective, although resorting to LTE links when located in a LTE cell, MMUEs will use direct messaging services over satellite in S-band when outside of AeNB or TeNB coverage. Moreover, S-band enabled mobile terminals shall be used as relays in both forward and return link directions to provide limited services to citizens in the disaster area where the regular terrestrial networks are disrupted: broadcast (alert and information) in the forward link and messaging in the return link (for example distress).

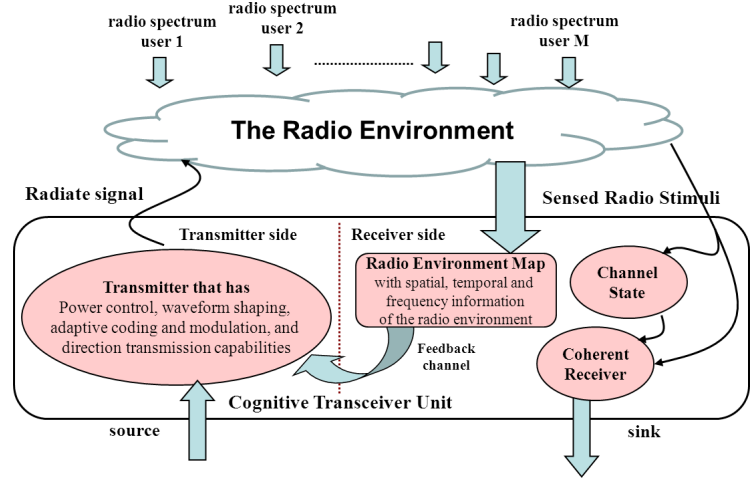

**Fig. 6.** Cognitive Cycle, as initially described by Mitola [15] and Haykin [16].

## 5 Cognitive Communications in PPDR

There, where the existing communications infrastructure is damaged or malfunctioning, scalable and adaptable network architectures are needed. Cognitive radio technology [15] is regarded here as a candidate to provide such flexibility in temporary deployments. In cognitive radio, wireless nodes learn about the radio environment and adapt their transmissions to maximize spectral efficiency. This is done by implementing the cognitive cycle (see Figure 6).

The cognitive cycle includes the mapping of the radio environment, in which nodes learn about and create a radio map of their surroundings. This allows the nodes to take intelligent decisions at the radio level such as Dynamic Spectrum Access and Energy Efficient Communications. The environmental radio mapping includes *spectrum sensing* and the *localization* of nearby radio users.

Though multiple *spectrum sensing* techniques exist, it is generally challenging because other radios need to be reliably detected. This is especially difficult when radio users are far from the sensing node or undetectable, thus leading to the 'hidden node problem'. To address this, cooperative spectrum sensing is considered, in which various cognitive radio nodes share the collected information. *Localization* is even more challenging since the radio to be located does not necessarily cooperate. Therefore no prior knowledge exists to find it satisfactorily. To cope with this, blind techniques have also been proposed [17].

### 5.1 Cooperative Spectrum Sensing

Let there be $K$ cognitive radio nodes, where each node senses alone and creates a test statistic $\xi_k$ with $k = 1, 2 \ldots K$ to detect the presence of a radio user. A local decision can be made based on a binary decision making criteria as,

$$d_k = \begin{cases} 1, & \text{if } \xi_k \geq \mu \\ 0, & \text{if } \xi_k < \mu \end{cases} \quad (3)$$

where, $\mu$ is the detection threshold. In cooperative sensing, the local decisions $d_k$ or the test statistics $\xi_k$ are sent to a central Cognitive Base Station (CBS), which collects the data and makes global decisions on behalf of the distributed radios and reports it to them. Through cooperative sensing the hidden terminal problem is greatly addressed. Let us consider next a cooperative sensing example based on local energy detection [18]. Let us assume the cognitive radio nodes send the local decisions $d_k$ to the CBS where $d_k$ are derived using energy test statistics $\xi_k$. If the CBS performs a fusion strategy $\Lambda(u_1, u_2 \ldots u_k, \ldots u_K)$ on the received data, and $u_k = d_k$, then the miss detection and false alarm probabilities for the corresponding detection mechanism are respectively given by,

$$P_{M,c} = \sum_{j=0}^{K-1} \frac{K!}{(K-j)!j!} P_D^j (1 - P_D)^{K-j} \quad (4)$$

$$P_{FA,c} = 1 - \sum_{j=0}^{K-1} \frac{K!}{(K-j)!j!} P_{FA}^j (1 - P_{FA})^{K-j}, \quad (5)$$

where $P_D$ and $P_{FA}$ are the detection and the false alarm probabilities for the local sensing performed at every cognitive radio node, assuming all the cognitive radio nodes have the same $P_D$ and $P_{FA}$ values, and $\Lambda(.)$ is a logical 'OR' function.

Figure 7 depicts the Receiver Operating Characteristic (ROC) curves of the sensing scheme described above for various numbers of cooperative nodes. This illustrates that the detection performance increases with the number of cooperating nodes. This is also compared with the non-cooperative scheme (i.e one single node or $K = 1$). ABSOLUTE will research cooperative sensing schemes taking into account the features of the various radios (e.g. sensors and UEs).

### 5.2 Dynamic Spectrum Access and Management

Based on the collected information about the radio environment, the nodes can utilize the radio spectrum dynamically and opportunistically by either reducing interference to other radios or by avoiding interference themselves. The spectrum access strategies adopted in the network can be tailored to the specific requirements set by the network policies. In essence two types of dynamic access are considered in this project, 1) underlay spectrum access where cognitive radios share the spectrum simultaneously whilst keeping interference to any radio below the levels assigned by the regulatory bodies, 2) overlay spectrum access where cognitive radios utilize the spectral gaps opportunistically to transmit.

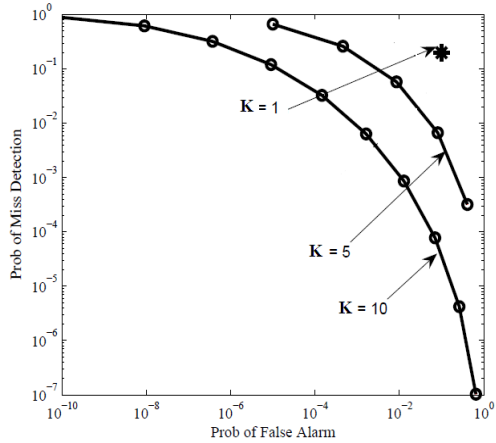

**Fig. 7.** Performance of a cooperative sensing scheme based on energy detection.

## 6 Conclusions

It has been highlighted, that during disasters or temporary events, communication networks play a supporting role and that new architectures are needed to provide flexible, scalable, resilient and secure broadband access.

Based on a rapidly deployable airborne platform with an embedded eNodeB, the ABSOLUTE system will quickly deploy mobile networks with large *local* coverage. This is thanks to the combination of multiple airborne, terrestrial and satellite systems. The system can be deployed to support first responders communications, to support the restoration of critical infrastructure and to serve as temporary (main or complementary) high-capacity communication infrastructure for large-scale events. The architecture described is well suited for public safety applications and it will also allow operators to increase their capacity during specific events such as the Olympic Games. In this context, the following challenges still remain to be solved in order to achieve the above listed goals:

– Development of a standalone eNodeB that can be embedded in an a platform. Factors such as weight, size, temperature range of the 4G components as well as the antenna need to be considered in these aeronautical systems.
– Cooperative spectrum sensing techniques that guarantee network resilience.
– Design of a satellite backhauling solution capable of providing the necessary capacity while keeping the system's mobility.

Moreover, as future work, we are currently evaluating the impact of LAP altitude on the network performance, in terms of capacity and delay, for several services running on parallel such as half-duplex video conferencing and LTE Push-to-talk. Besides, we are also evaluating the most suitable techniques in order to optimize the ABSOLUTE system capacity for the network infrastructure deployed for supporting disaster-relief activities and extending capacity during temporary mass events.

# 7 Acknowledgement

This document has been produced in the context of the ABSOLUTE project. ABSOLUTE consortium wants to acknowledge that the research leading to these results has received funding from the European Commissions Seventh Framework Programme (FP7-2011-8) under the Grant Agreement FP7-ICT-318632.